\documentclass[aps,physrev,superscriptaddress,twocolumn,showkeys,floatfix]{revtex4-2}

\usepackage{graphicx,import}
\usepackage{dcolumn}
\usepackage{bm}

\usepackage{placeins}
\usepackage[version=4]{mhchem}
\usepackage{siunitx}
\usepackage{xcolor}
\usepackage{float}
\usepackage{ulem}

\footnotetext{These authors contributed equally to this work.}

\begin{document}


\title{\textbf{Second harmonic study of thermally oxidized mono- and few-layer 2H-\ce{MoS2}} 
}%


\author{Katharina Burgholzer$^{\ddagger}$}
\email{katharina.burgholzer@jku.at}
\affiliation{Institute of Semiconductor and Solid State Physics, Johannes Kepler University Linz, 4040 Linz, Austria}

\author{Henry Volker Hübschmann$^{\ddagger}$}
\email{henry.volker.huebschmann@uni-paderborn.de}
    \affiliation{Department of Physics, Paderborn University, 33095 Paderborn, Germany}
    \affiliation{Institute for Photonic Quantum Systems (PhoQS), Paderborn University, 33095 Paderborn, Germany}
    \affiliation{Center for Optoelectronics and Photonics Paderborn (CeOPP), Paderborn University, 33095 Paderborn, Germany}

\author{Gerhard Berth}
    \affiliation{Department of Physics, Paderborn University, 33095 Paderborn, Germany}
    \affiliation{Institute for Photonic Quantum Systems (PhoQS), Paderborn University, 33095 Paderborn, Germany}
    \affiliation{Center for Optoelectronics and Photonics Paderborn (CeOPP), Paderborn University, 33095 Paderborn, Germany}

\author{Adriana Bocchini}
    \affiliation{Department of Physics, Paderborn University, 33095 Paderborn, Germany}
    \affiliation{Institute for Photonic Quantum Systems (PhoQS), Paderborn University, 33095 Paderborn, Germany}
    \affiliation{Center for Optoelectronics and Photonics Paderborn (CeOPP), Paderborn University, 33095 Paderborn, Germany}
    
\author{Uwe Gerstmann}	 
    \affiliation{Department of Physics, Paderborn University, 33095 Paderborn, Germany}
    \affiliation{Center for Optoelectronics and Photonics Paderborn (CeOPP), Paderborn University, 33095 Paderborn, Germany}
    
\author{Wolf Gero Schmidt}
    \affiliation{Department of Physics, Paderborn University, 33095 Paderborn, Germany}
    \affiliation{Center for Optoelectronics and Photonics Paderborn (CeOPP), Paderborn University, 33095 Paderborn, Germany}

\author{Klaus D. Jöns}
    \affiliation{Department of Physics, Paderborn University, 33095 Paderborn, Germany}
    \affiliation{Institute for Photonic Quantum Systems (PhoQS), Paderborn University, 33095 Paderborn, Germany}
    \affiliation{Center for Optoelectronics and Photonics Paderborn (CeOPP), Paderborn University, 33095 Paderborn, Germany}

\author{Alberta Bonanni}
\affiliation{Institute of Semiconductor and Solid State Physics, Johannes Kepler University Linz, 4040 Linz, Austria}


\date{\today}

\begin{abstract}

A comprehensive study of second harmonic generation on thermally oxidized \ce{MoS2} flakes with thickness ranging from monolayer up to seven layers is presented. Observing the fundamental nonlinear behavior for non-treated and oxidized \ce{MoS2} reveals that oxidation causes significant changes in the second harmonic (SH) response for all investigated structures. Excitation power dependent measurements to analyze the nonlinear behavior with respect to the oxidation time show progressive oxidation within the maximum oxidation time of six hours, under the considered oxidation conditions. Here, polarization dependent measurements reveal the structural changes due to oxidation. Additionally, it is found that the oxidation depth is restricted to the top most layer and the oxidation behavior exhibits a layer dependency. These findings are supported by theoretical band structure calculations. The results demonstrate that the thermal oxidation progress of two dimensional \ce{MoS2} can be monitored with non-resonant and non-invasive SH microscopy, by following distinct fingerprints of structural modification in the nonlinear response.

\end{abstract}


\maketitle


\section{Introduction}

Two dimensional (2D) semiconductors belonging to the family of transition metal dichalcogenides, have attracted increasing attention as building blocks for next generation nanoelectronics. In particular, molybdenum disulfide (\ce{MoS2}), besides showing an electron mobility up to \SI{200}{\centi\metre\squared\per\volt\per\second} \cite{Radisavljevic2011Single-layerTransistors}, it also has a thickness-dependent band structure and band gap ranging from indirect of \SI{1,2}{eV} in the bulk state \cite{Kam1982DetailedDichalcogenides} to direct of \SI{1,8}{eV} in monolayer form \cite{Mak2010AtomicallySemiconductor} at room temperature. For this reason, \ce{MoS2} is already applied in atomically thin devices for nanoelectronics \cite{Radisavljevic2011Single-layerTransistors, Wang2018RobustMaterials}, photovoltaics and photodetection \cite{Amani2015Near-unitysub2/sub, Yin2012Single-layerPhototransistors}, chemical sensorics \cite{Li2012FabricationTemperature, Perkins2013ChemicalMoS2}, and electrochemical catalysis \cite{Wang2018ControllableSites}. Thermal oxidation is involved in the fabrication of such devices in order to tune their electronic and optical properties \cite{Wang2018RobustMaterials}. A comprehensive understanding of the thermal oxidation mechanism of \ce{MoS2} is crucial to ensure the reliability and optimized performance of \ce{MoS2}-based devices. Since an odd number of 2H-\ce{MoS2} layers exhibits no inversion symmetry and gives rise to second-order optical nonlinearity ($\chi^{(2)}\not=0$) and the structural impact by oxidation is expected, second harmonic (SH) microscopy a helpful tool for analyzing the crystal structure and changes due to oxidation. Here, a comprehensive study is conducted on how oxidation affects the nonlinear optical response of 2H-\ce{MoS2} flakes, with thickness ranging from 1 layer (L) up to 7L on \ce{SiO2}.

\section{Material system}
\subsection{Crystal symmetry}

\begin{figure}[!hhh]
\centering
\includegraphics[width=0.35\textwidth]{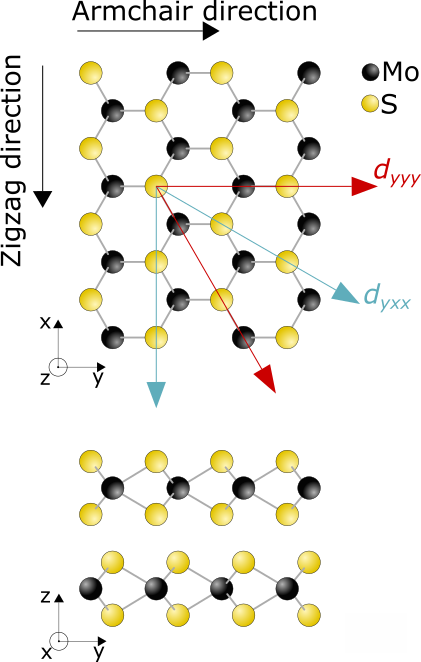}
    \caption{Crystal structure of 2H-\ce{MoS2} (top) top view with indications of the armchair and zigzag directions and corresponding orientation of the nonlinear tensor elements $d_{yyy}$ and $d_{yxx}$, (bottom) side view 2H stacking.}
    \label{fig:skizze}
\end{figure}

A monolayer (1L) \ce{MoS2} contains two hexagonal sulfur (S) lattices with a layer of molybdenum (Mo) atoms positioned in trigonal prismatic coordination between the S planes, leading to the crystal structure with a broken inversion symmetry, shown in Figure \ref{fig:skizze}.
The 1L \ce{MoS2} belongs to the D$_{3h}$ symmetry group, with only one independent, non-vanishing nonlinear tensor element $d_{yyy}=- d_{yxx}=-d_{xxy}=-d_{xyx}$ along the high symmetry directions, armchair and zigzag, each with a three-fold symmetry. 
For a bilayer (2L) 2H-\ce{MoS2} another S-Mo-S layer is added with AA'-stacking \cite{He2014Stacking2}, resulting in a centrosymmetric D$_{3d}$ space group. Consequently, for an odd number of layers the inversion symmetry is broken, whereas for an even number the inversion symmetry is restored \cite{Molina-Sanchez2015VibrationalBulk}.

\subsection{Sample fabrication}

The investigated MoS$_{2}$ samples are fabricated by mechanical exfoliation \cite{Novoselov2005} from a bulk 2H-MoS$_{2}$ crystal obtained from MaTeck. 
For the first inspection and the further transfer step, the MoS$_{2}$ flakes are placed onto viscoelastic polydimethylsiloxane (PDMS) stamps with a rigid metallic support. Suitable MoS$_{2}$ flakes are transferred onto Si substrates covered with \SI{285}{nm} of SiO$_{2}$ using a custom deterministic all-dry transfer setup. To avoid a full oxidation, preliminary tests at various temperatures have been carried out and the specimens have been monitored under an optical microscope and by X-ray photoemission spectroscopy. These tests show that \SI{300}{\degreeCelsius} is the temperature of choice for a visible oxidation of the chosen material configuration, without destroying the crystal structure, what is in agreement with previous studies \cite{Jiang2023PhysicalBeyond}. Therefore, in a last fabrication step, the samples are heated to $T=$ \SI{300}{\degreeCelsius} with a ramp time of \SI{200}{s} and annealed for different times $t_a=$(0, 2, 4, 6) hours in a MILA-5000 infrared lamp heating system with controlled oxygen (O) flow of 15 standard cubic centimeters per minute (sccm) throughout the thermal oxidation process.

\subsection{Pre-characterization}

The determination of the shape and thickness of few-layer MoS$_{2}$ is obtained with a high-resolution Keyence VHX-7000 optical microscope. In the optical images structural defects, like \textit{e.g.}, wrinkles, are visible. Furthermore, the contact between the flake and the substrate can be ensured. The number of \ce{MoS2} layers up to 9L can be identified by comparison of the contrast and color \cite{Castellanos-Gomez2010OpticalCrystals}.
\begin{figure}[!hhh]
\includegraphics[width=0.5\textwidth]{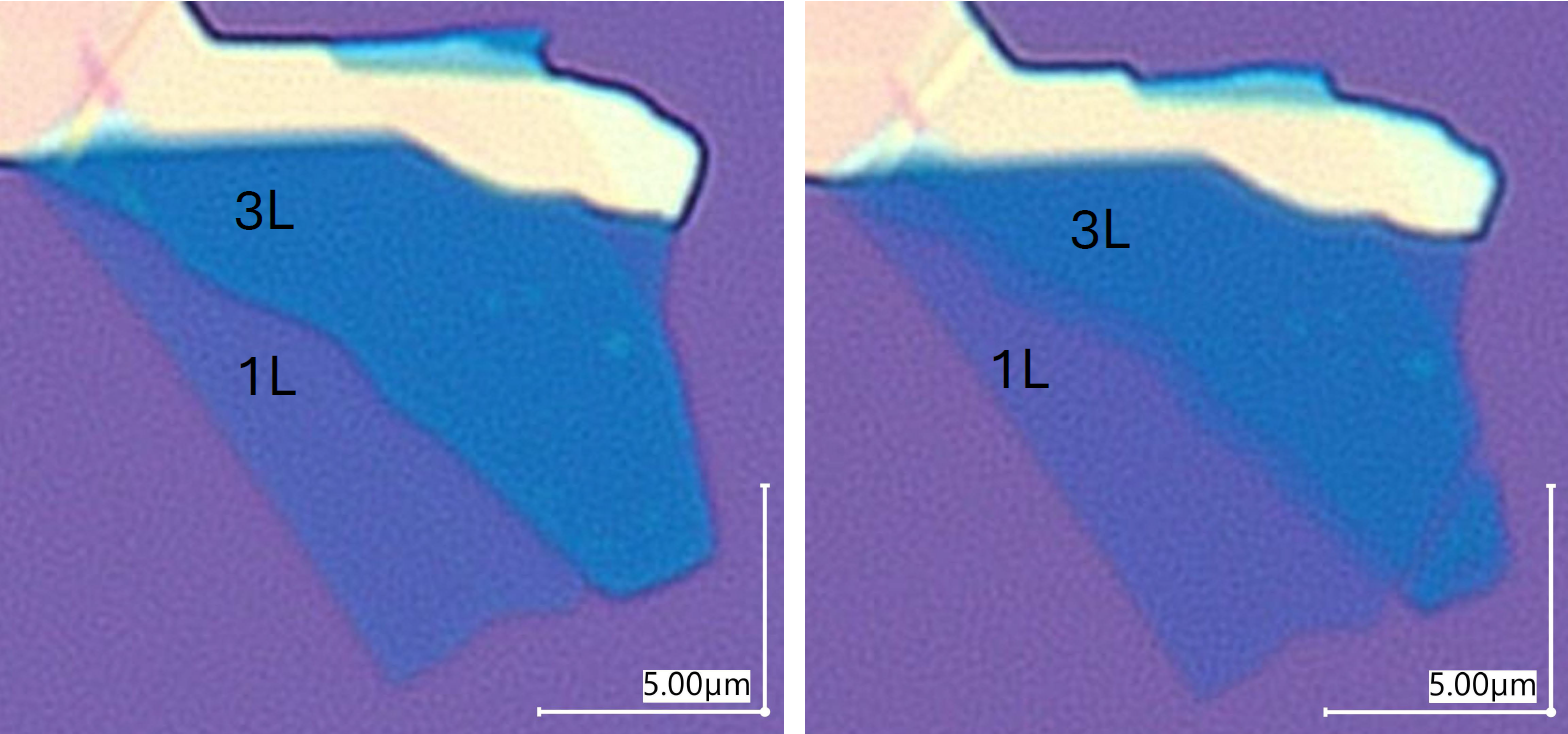}
    \caption{Optical images of \ce{MoS2} flakes on \ce{SiO2}/Si-substrate before (left) and after (right) 6h-oxidation at \SI{300}{\degreeCelsius}. Here, the regions for the 1L and 3L system are labeled. Structural modifications due to oxidation become visible in a slight change of contrast. Especially at the edges of the regions of different thicknesses a significant modification is observed.}
    \label{fig:optical}
\end{figure}
All structures have been inspected before and after oxidation, with respect to  macroscopic structural modifications induced by annealing and O exposure. Exemplary images of one sample with a 1L and a 3L region before and after six-hour oxidation are shown in Figure \ref{fig:optical}. The thermal oxidation treatment of \ce{MoS2} flakes induces a slight change of contrast, as evidenced by the comparison between the 1L and 3L area in Figure \ref{fig:optical}. Furthermore, partially damaged areas and highly oxidized \ce{MoS2} can be discerned. Especially at the edges of the regions of different thicknesses and at the defects, a significant modification is observed, likely due to dangling bonds at the edges \cite{Jiang2023PhysicalBeyond}. For the further analysis, only regions which appear homogeneous in the optical images are considered.

\begin{figure}[htpb]
\includegraphics[width=0.44\textwidth]{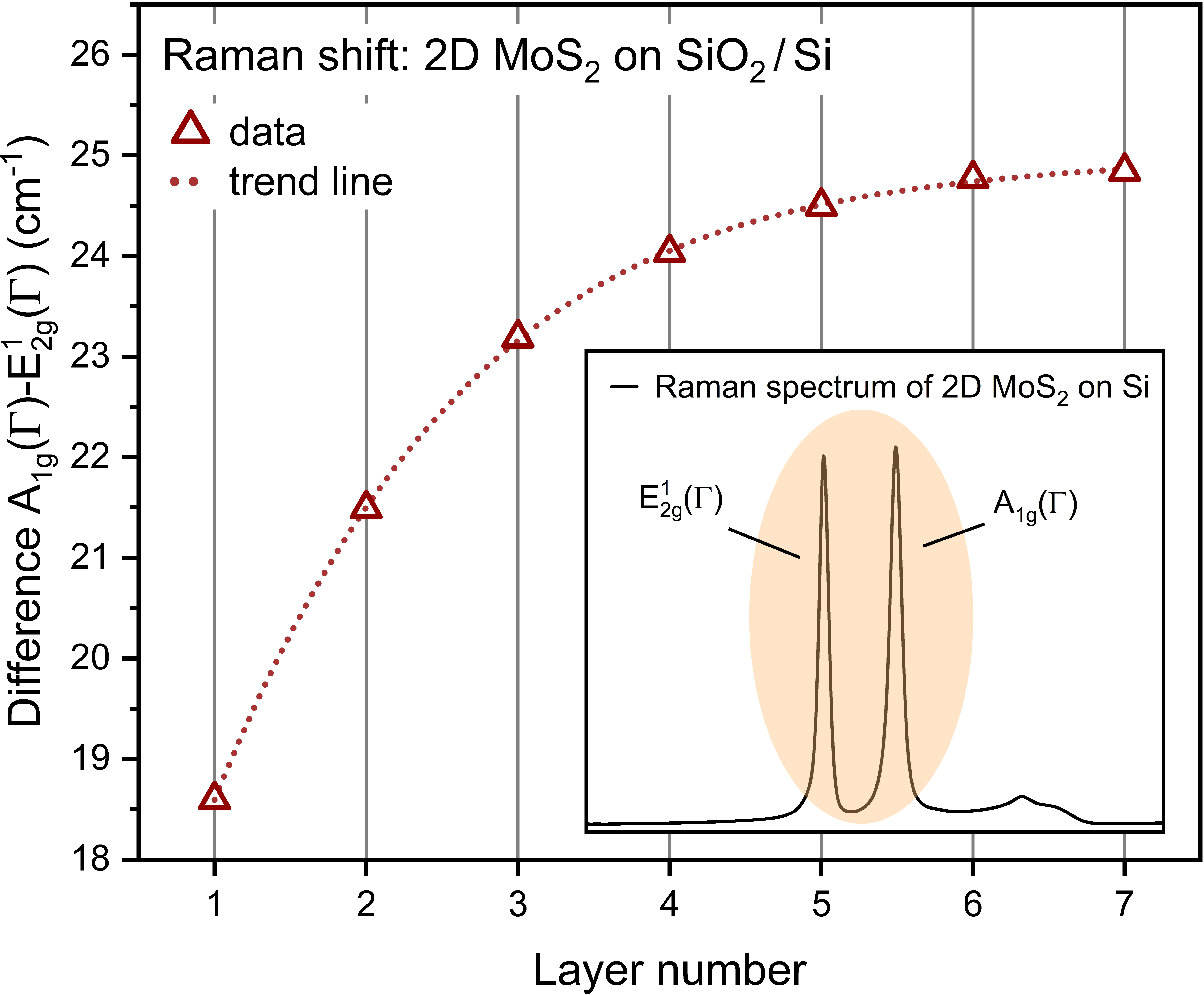}
    \caption{Absolute Raman shift difference progression with layer number for \ce{MoS2} on \ce{SiO2}/Si-substrate. The shift difference is taken between the phonon modes E$^{1}_{2g}(\Gamma)$ and A$_{1g}(\Gamma)$, which undergo a red- (blue-) shift with increasing layer number, respectively. 
    }
    \label{fig:raman}
\end{figure}

The individual layer number dependent optical contrast of the untreated \ce{MoS2} flakes is calibrated \textit{via} Raman spectroscopy \cite{Lee2010AnomalousMoS2}. The determination of few-layer numbers relies primarily on the evolution of the Raman shift of the two dominant Raman modes E$_{2g}^{1}$ and A$_{1g}$, which are red- and blue-shifted, respectively, with increasing layer number. To avoid systematic disturbances, the more stable absolute Raman shift difference between both phonon modes is considered, plotted in Figure \ref{fig:raman} as a function of the layer number. Due to the convergence to their bulk Raman shift of each phonon mode \cite{Lee2010AnomalousMoS2}, the absolute Raman shift difference also converges for increasing layer numbers. Thus, Raman analysis is reliable for few-layer numbers up to 5L. For a higher layer number, additional information is required in order to determine the thickness, \textit{e.g.}, whether there is an even or odd number of layers, which can be evidenced from the SH-intensity, as to be discussed later. In this work, the analyzed non-oxidized and oxidized samples thickness range from a 1L to 7L. 

\section{Methods}
\subsection{Experimental environment}

The nonlinear analysis presented in this work has been conducted with a custom-build nonlinear optical setup, which allows for power- and polarization-controlled measurements. The excitation light from a Toptica FemtoFiber Smart 780 pulsed laser (central wavelength \SI{783}{nm} ± \SI{5}{nm}, \SI{100}{fs} pulse duration, \SI{80}{MHz} repetition rate) is focused onto the sample through an infinite corrected objective (NA = 0.95, 100x mag.). The generated SH-light is collected by the same objective in backscattering direction and detected by a single photon counting module (silicon APD). For spectral filtering (to avoid detection of excitation light), a dichroic beamsplitter and appropriate color filters are implemented. In the framework of a fixed laserfocus the sample positioning is realized by a 3D positioning unit.
Due to the unknown orientation of the MoS$_2$ flakes with respect to the laboratory frame, the identification of the individual crystal axes is essential. To obtain full orientation information, the incident laser polarization is defined \textit{via} a half-wave plate and the detection polarization is specified by using a pol-analyzer. A simultaneous rotation of both allows for scanning the six-fold symmetry of one specific tensor element. If the polarization of the incident laser is perpendicular to the detection polarization, the zigzag-direction corresponding to $d_{yxx}$ is detected, and if they are parallel aligned, the armchair-direction $d_{yyy}$ is probed. In this work, the acquisition of the SH-intensity is performed in parallel polarization orientation in alignment with the corresponding crystal axis, due to the equivalence of the absolute value of both tensor elements.

\subsection{Computational environment}

Ground-state crystal geometries and electronic structures are obtained within density functional theory (DFT), as implemented in the open-source program package \textsc{Quantum ESPRESSO} \cite{Giannozzi2009, Giannozzi2017}.
The spin-orbit coupling (SOC) is neglected in the calculations and only the fully oxidized case is considered, \textit{i.e.}, the top S layer is fully replaced by O atoms.

To model the oxidized MoO$_x$S$_{2-x}$ few-layer systems under periodic boundary conditions, the lattice parameters $a=b=3.163$\,\AA\ are kept fixed at the experimental value \cite{fan2014} of MoS$_2$ and about $10$\AA\ of vacuum are added along the $[001]$ crystal direction on up to 3L of MoS$_2$. 
The atomic positions, on the other hand, are relaxed until the fluctuations of forces and the total energy are below $10^{-8}$\,Ry/Bohr and $10^{-4}$\,Ry, respectively.
In addition, plane waves are expanded up to an energy cutoff of $90$\,Ry and the Brillouin zone is sampled using a $\Gamma$-centered $12 \times 12 \times 1$ \textbf{\textit{k}}-point mesh.

The electronic exchange and correlation (XC) effects are modeled within the generalized gradient approximation using the Perdew-Burke-Ernzerhof functional (PBE) \cite{pbe}.
In addition, norm-conserving pseudopotentials are employed and the van der Waals interactions are accounted \textit{via} the semi-empirical Grimme D3 scheme \cite{grimmed3}.

\section{Nonlinear analysis}

\subsection{Fundamental behavior}

\begin{figure}[htbp]
\includegraphics[width=0.45\textwidth]{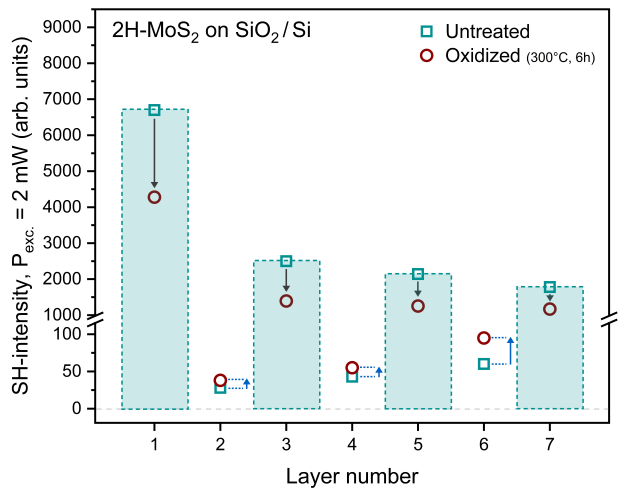}
    \caption{Absolute SH-signal for untreated and oxidized 2H-\ce{MoS2} samples with respect to the layer number. Changes due to oxidation are indicates \textit{via} arrows. Measurements are performed @ \SI{2}{mW} laser power and \SI{1}{s} integration time.}
    \label{fig:generalSH}
\end{figure}

To study the influence of thermal oxidation on the nonlinear optical behavior of the samples, a layer dependent analysis has been performed. Specifically, untreated and oxidized ($6$h @ \SI{300}{\degreeCelsius}) samples are probed and the corresponding SH-signal for different layer configurations is compared, as reported in Figure \ref{fig:generalSH}, where the SH-intensity as a function of layer number is displayed for the untreated and for the oxidized case.

In the untreated case, a nonlinear signal for all configurations with odd layer number parity is observed. This is due the lack of inversion symmetry, enabling second harmonic generation (SHG). Furthermore, it becomes apparent that for the odd layer configurations, the SH-signal decreases with increasing layer number in a nonlinear manner. This behavior is assigned to the increase of absorption and changes in the band structure with increasing sample thickness ($E_{\mathrm{SHG}}>E_{\mathrm{gap}}$ for 2L and more) \cite{Castellanos2016Nano} \cite{shgmos22013}, here resulting in a quenched light-matter interaction \cite{SplendianiNanoL}.

\begin{figure}[htbp]
\includegraphics[width=0.35\textwidth]{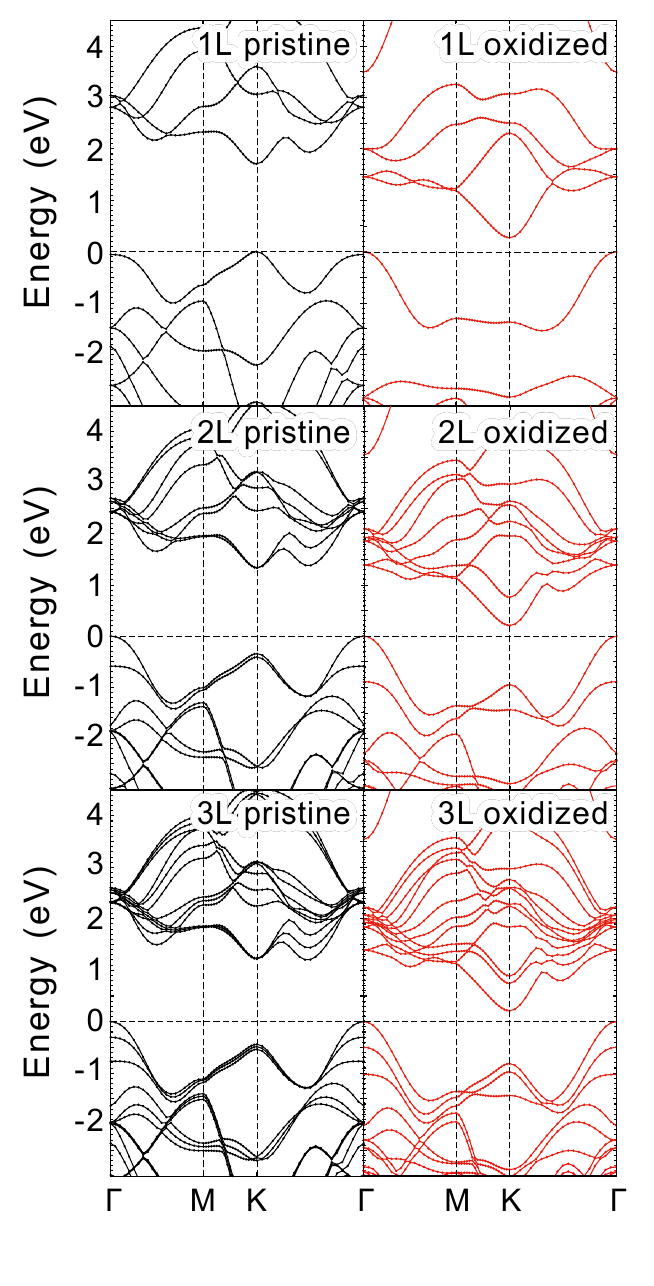}
    \caption{DFT band structure of 1L -3L for pristine \ce{MoS2} and the influence of full exchange of the top S layer with O atoms (oxidized).
    With a coupling constant of  $\lambda = 75$ meV \cite{zhu2011, nguyen2018}, the effect of SOC is small in comparison with the changes introduced by oxidization and thus can be safely neglected here.}
    \label{fig:bandstructure}
\end{figure}

For even-numbered layer configurations, no SH-light is expected, due to the inversion symmetry of the 2H-AA'-stacked systems \cite{Li2013ProbingGeneration}. Nevertheless, a residual signal (orders of magnitude lower compared to the one from the odd numbered configurations) is observed, which increases with increasing layer numbers. The signal is likely to originate from interlayer-effects \cite{ShreeNatureC} or from changing interlayer bonding lengths caused by the substrate and resulting in a break of symmetry \cite{mos2substrate}. An inherent asymmetry therefore leads to an increase of the observed SH-signal for increasing layer numbers, as more material is available for SHG.

Compared to the untreated case, oxidation causes significant changes in the SH response for all layered structures. In case of even-layer structure, the perfect inversion symmetry is broken by the higher amount of O in the topmost layer, so that a
relevant SH signal is introduced in the previously SHG-silent
structures \cite{etching2013}.
The case of odd-layer structures, however, is more involved and
cannot	be explained by	simple symmetry arguments alone.
We observe a reduction to about 50 to 70\% for every number of layers, suggesting a qualitative change
(moderate reduction) of the SHG (i) by the lighter, not so easy
to polarize O atoms, (ii) due to clusters with local inversion
symmetry and/or (iii) by oxygen induced changes in the electronic
structure.

\begin{figure*}[htbp]
\textbf{a)}
\includegraphics[height=4.3cm]{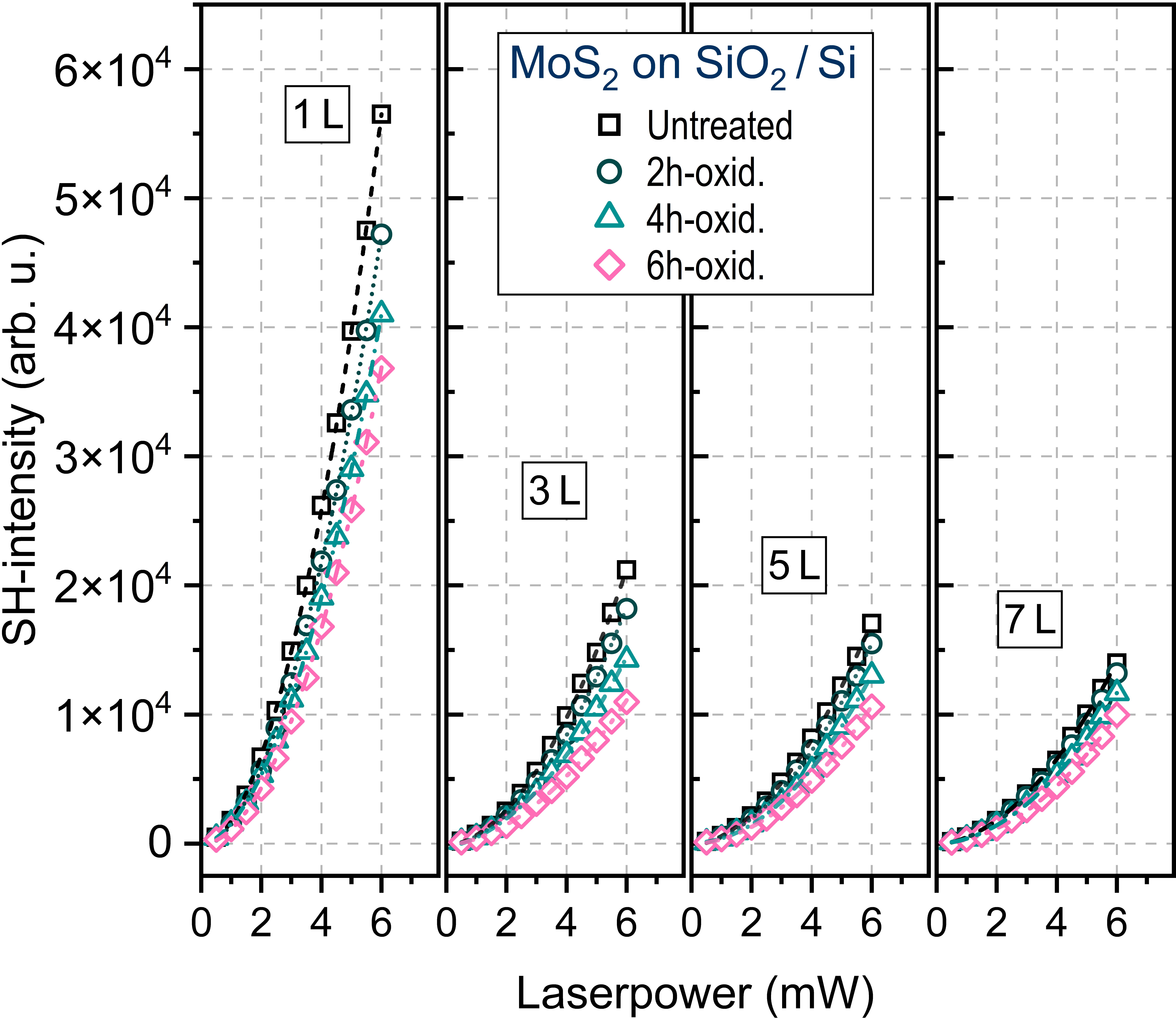}
\hspace{2mm}
\textbf{b)}
\includegraphics[height=4.3cm]{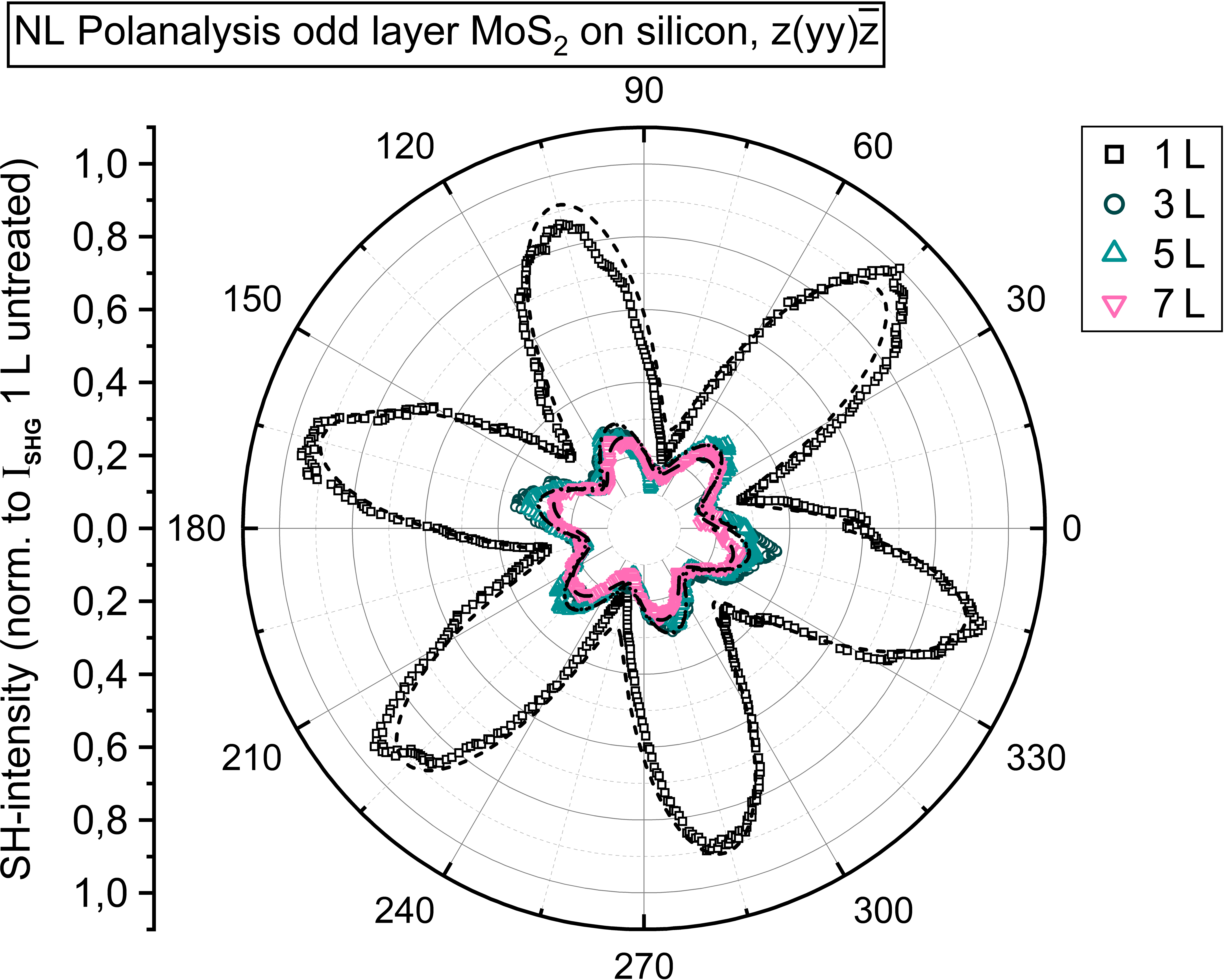}
\textbf{c)}
\includegraphics[height=4.3cm]{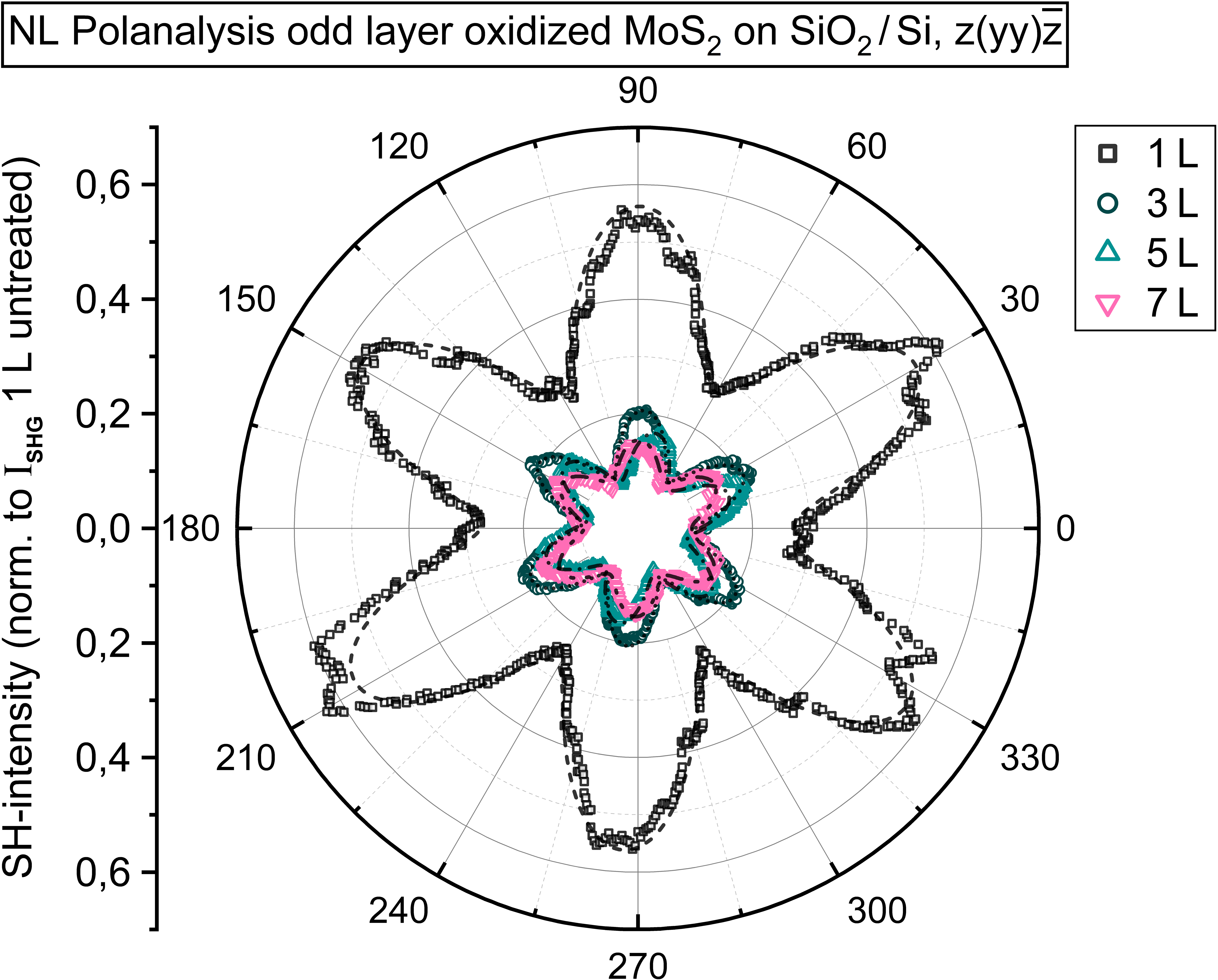}
    \caption{a) Power dependent measurements showing the progression of the SH-intensities with laser power for the odd layer configurations. The measurements are separated for each layer with corresponding SH-intensity progression after 0, 2, 4 and 6h oxidation time. Polarization dependent nonlinear analysis of odd layered 2D \ce{MoS2} before (b) and after (c) \SI{6}{h} oxidation. The observed signal is normalized to the maximum intensity from 1L. The data is obtained for \SI{1}{mW} laser power with angular resolution of \SI{0.5}{degree}.}
    \label{fig:PDoxidodd}
\end{figure*}

For a better understanding of the signal change induced by oxidation, the band structures of 1L -3L of \ce{MoS2} are calculated, as plotted in Figure \ref{fig:bandstructure}, where the band structure (left) of the pristine case is compared with (right) a fully oxidized top S layer (top-most S layer exchanges against O).
For a 1L structure the oxygen-induced change in the band structure is most pronounced: 
The bands at the K point hosting the direct band gap for a pristine MoS$_2$ monolayer are downward-shifted by more than \SI{1}{eV}.
A similar shift is observed at the $\Gamma$ point, but here the shift
is restricted to the conduction bands, resulting  in a transition
from direct to an indirect band gap. 
By this, the fundamental gap is not only reduced to less than \SI{0.5}{eV}. Interestingly, also the curvature of the topmost valence band between the M and K points and in particular the lowest conduction bands around the $\Gamma$ point is affected and essentially flattened. 
As a result the previously characteristic identical dispersion of VBM and CBM with energy difference of about \SI{3.16}{eV} (corresponding to  $2 \times 1.58$\,{eV} photon energy) is lifted, suggesting a considerable decrease in the combined density of states, providing a natural explanation of the observed O-induced decrease in SHG intensity.
In other words, the nonlinear response of odd number of layers is not
necessarily weakened due to oxidation in general, but presumably
shifted spectrally and strongly influenced by band bending, which
expresses itself in a decreased SHG signal at the experimentally
used center wavelength of \SI{783}{nm} (\SI{1.58}{eV} photon energy).

\subsection{Power and polarization dependency}

In a next step the nonlinear behavior with respect to oxidation time $t_a=(0, 2, 4, 6) $h is analyzed in order to proof the progress of oxidation. Here, excitation power dependent ($0-6$)mW measurements are performed for each layer configuration. In order to corroborate the power dependent results, comparative polarization resolved measurements in armchair direction ($d_{yyy}$) are carried out on untreated and 6h oxidized \ce{MoS2} for up to 7L. For all polarization resolved measurements, an excitation power of \SI{1}{mW} is applied to prevent laser induced structural modifications. The fundamentally different nonlinear behavior of even and odd number of layers is discussed separately, starting with odd layers.

\begin{figure*}[htbp]
\textbf{a)}
\includegraphics[ height=4.3cm]{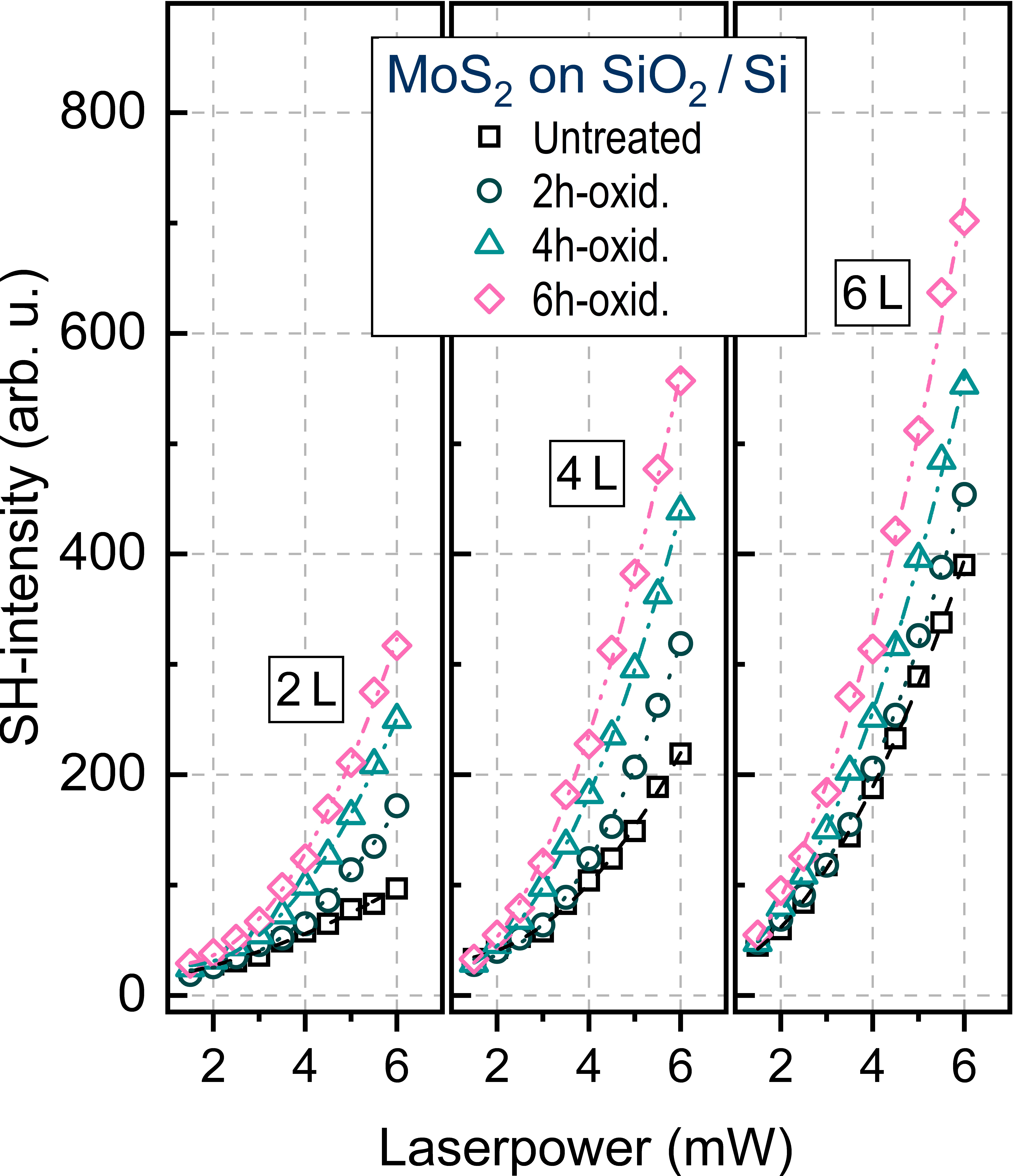}
\textbf{b)}
\hspace{2mm}
\includegraphics[ height=4.3cm]{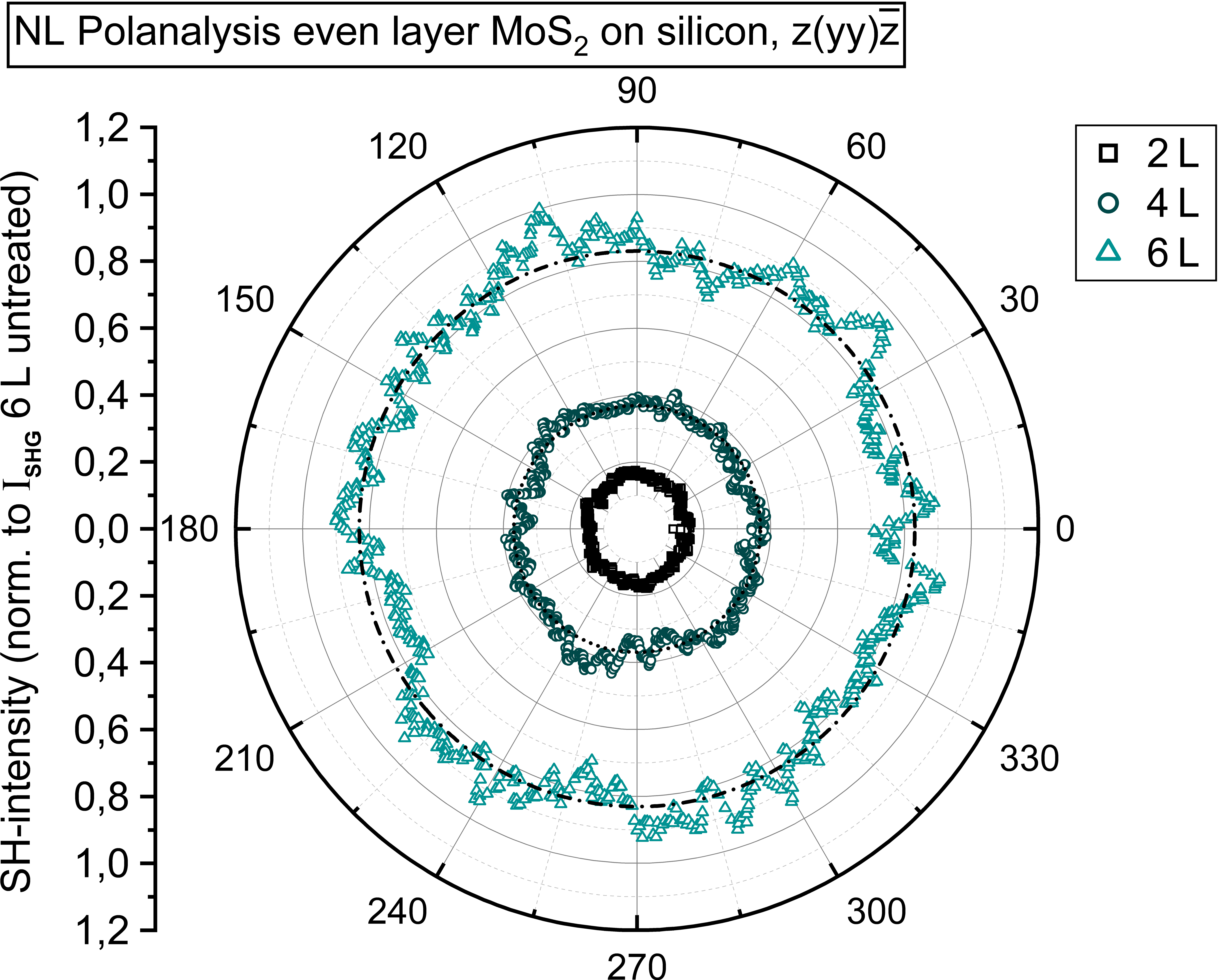}
\textbf{c)}
\includegraphics[ height=4.3cm]{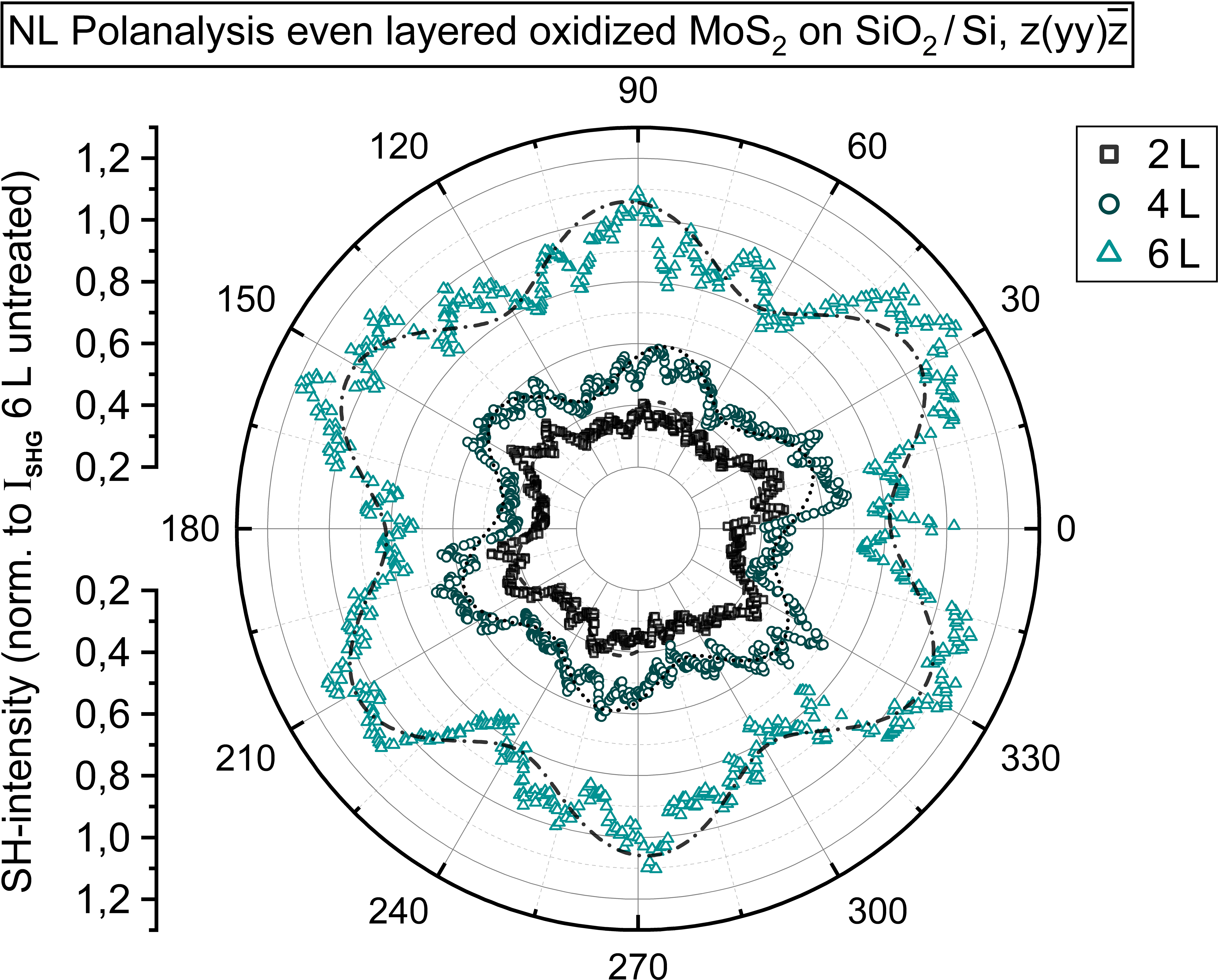}
    \caption{a) Progression of the SH-intensities with laser power for even layered configurations with corresponding SH-intensity evolution after 0, 2, 4 and 6h oxidation time (parabolic fits: $f(x)=a+b\cdot x+c\cdot x^{2}$). Polarization dependent nonlinear analysis of even layered 2D \ce{MoS2} before (b) and after (c) \SI{6}{h} oxidation. The observed signal is normalized to the maximum intensity from 6L. The data is obtained for \SI{1}{mW} laser power with angular resolution of \SI{0.5}{degree}.}
    \label{fig:PDoxideven}
\end{figure*}

As already observed in the fundamental behavior, the SH-intensity decreases with increasing layer numbers for odd numbered configurations. This behavior is found for every individual level of oxidation, as plotted in Figure \ref{fig:PDoxidodd} a), where the progression of the SH-intensity with laser power for odd numbered 1L -7L MoS$_2$ is shown. Furthermore, the SH-intensity decreases consecutively with rising oxidation time for each layer configuration. This indicates a progressive oxidation within the maximum oxidation time of six hours, but there is a strong SH-signal after every oxidation time. In fact, similar studies on 1L and 2L have been performed where the second harmonic response was probed for different degrees of oxidation but using oxygen plasma \cite{Ko2016}. Here, a distinct layer-by-layer conversion of \ce{MoS2} to \ce{MoO3} was observed. The signatures of full layer oxidation are a lost SH-signal from the 1L and a strong signal from the 2L, comparable to the pristine \ce{MoS2}-1L due to introduced lack of inversion symmetry. Although we observe a strong change in the nonlinear response after oxidation, the magnitude of the change is not in the expected range for a full change of symmetry conditions. Therefore, we predict an oxidation depth no deeper than the first sulfur layer for our thermal oxidation treatment.
Additionally, the absolute decrease of SH-intensity due to oxidation reduces for increasing layer numbers.
In the polarization dependent SH-intensity, a six-fold dependency arises, as shown in Figure \ref{fig:PDoxidodd}, where the SH-intensity before (b) and after 6h oxidation (c) as a function of the polarization direction is displayed. After oxidation, the six-fold geometry is retained, including the respective reduction of the nonlinear signal. Furthermore, the difference between maximum and minimum intensity decreases in the oxidized case, which indicates that the oxidation has a specific preferred direction, also possibly dependent on the number of layers \cite{zhou2013thickness}.

The power dependent SHG analysis for even numbered configurations, shown in Figure \ref{fig:PDoxideven} a), exhibits behavior opposite to that of odd layers induced by oxidation. For each untreated configuration, the SH-intensity increases with incident laser power. The oxidation leads to an increase of the SH-response with treatment time for every layer configuration, pointing to an ongoing oxidation process. Compared to corresponding odd numbered configurations (\textit{e.g.,} 6L to 5L) the resulting SH-signal after a six hour oxidation for an even numbered configuration is still an order of magnitude lower. Thus again a full oxidation of the top-most layer can be excluded and oxidation is confined to the surface region. In addition, each individual SH-signal for a specific oxidation level (0, 2, 4 and 6 h @ specific laser power) increases with increasing layer number, whereby for all oxidation parameters (including the untreated case) the evolution of the signal occurs in a nonlinear manner.

In Figure \ref{fig:PDoxideven}, the SH-intensity for even numbered layers is shown before (b) and after 6h oxidation (c) as a function of the polarization direction. Non-oxidized even numbered layers give no angle dependent polarization, in accordance with the inversion symmetry. Only the residual signal is detected. Upon oxidation, a sixfold symmetry of the induced polarization is found. The six-fold polarization dependency is inherent proof for the breaking of inversion symmetry in the system, as induced by the oxidation treatment and therefore supports the previously performed power dependent analysis. It is worth emphasizing that the polarization dependency before and after oxidation is measured at the same laser power, therefore the six-fold pattern can only originate from the O treatment.

\subsection{Influence of oxidation time}

\begin{figure*}[htbp]
    \textbf{a)}
    \includegraphics[width=0.45\textwidth]{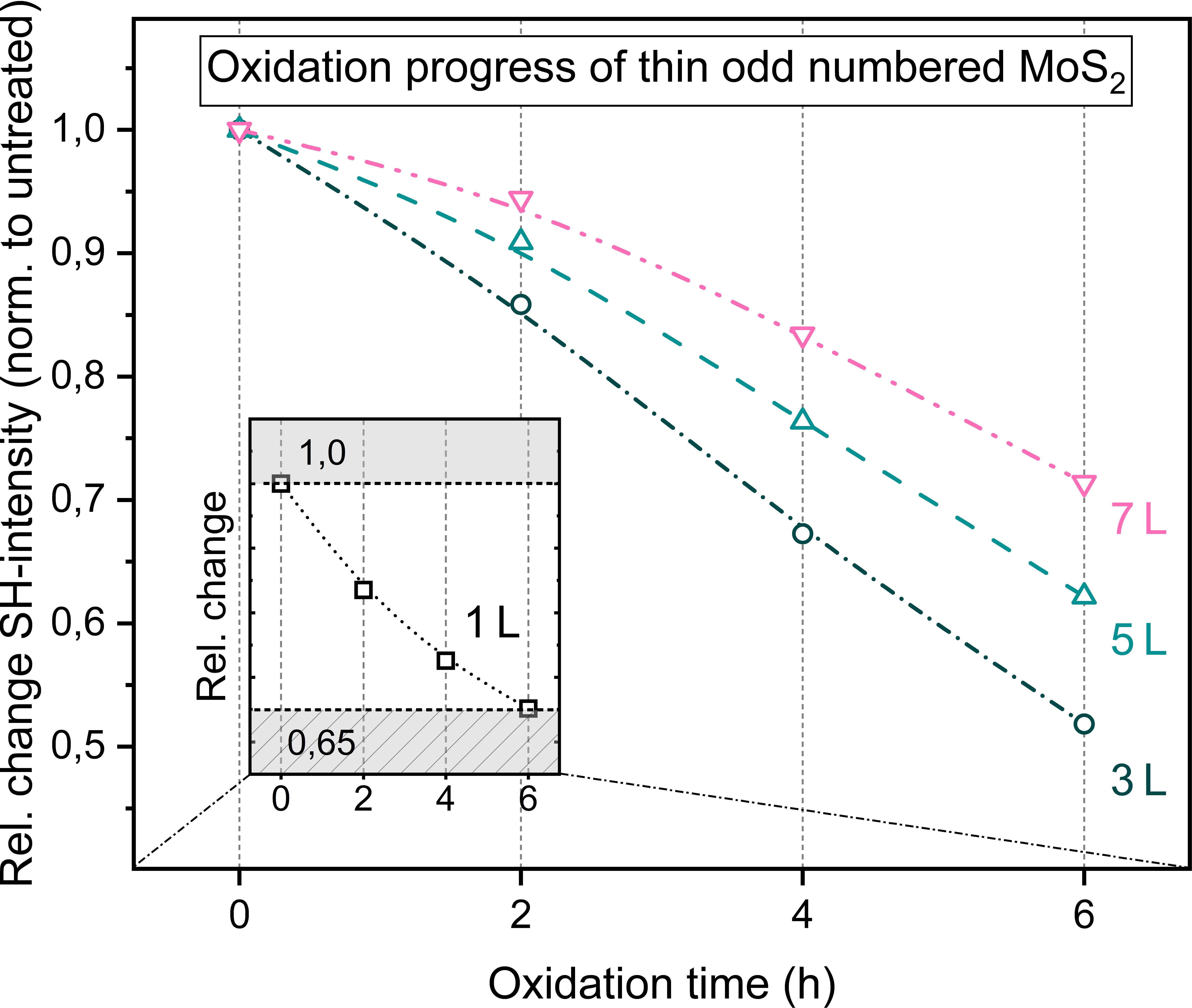}
    \textbf{b)}
    \includegraphics[width=0.45\textwidth]{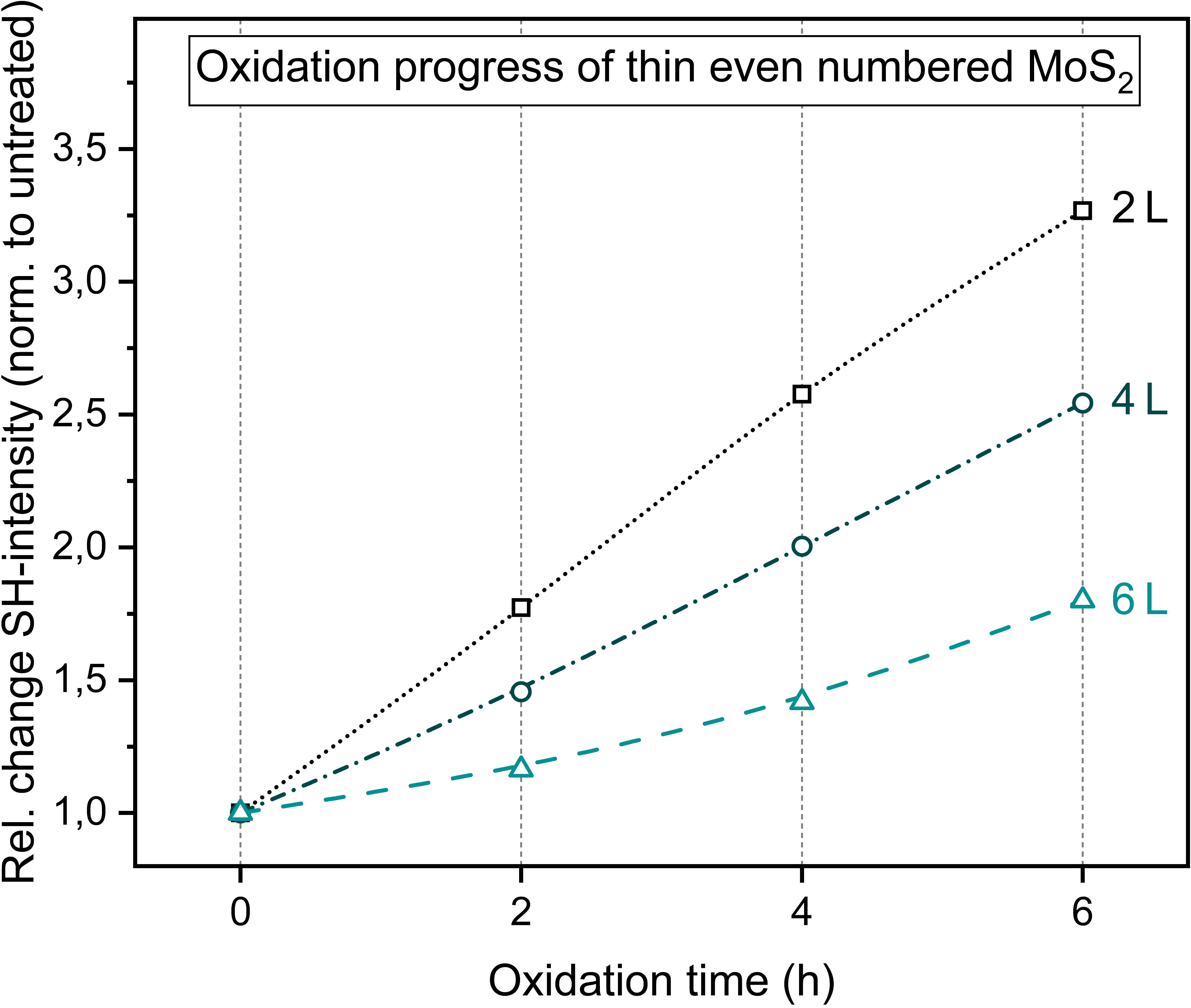}
  \caption{Oxidation induced relative change of the SH-intensity normalized to untreated case (laser excitation @ \SI{2}{mW}), with respect to the treatment time for odd (a) and even (b) layered configurations. Due to its fundamental different evolution the 1L is individually plotted in the inset to (a).}
  \label{fig:reloxidation}
\end{figure*}

In order to gain a deeper insight into the oxidation process, the relative change of SH-intensity for each number of layers is considered. Although the change in the absolute SH-signal for even and odd layers is quite different, the relative change in SH-intensity is greater the lower the number of layers, for both sets of layer number parity. This specific layer dependent effect is highlighted in Figure \ref{fig:reloxidation}, where the oxidation time dependent relative change of the SH-intensity is normalized to the untreated case and depicted for every odd (a) and even (b) layered configurations. 

The relative SH-signal at a fixed laser power with increasing oxidation time is approximately linear for each layer number, with the exception of the 1L, which is a special case to be discussed later. The increasing binding energies with increasing layer number \cite{Lin2016Thickness-DependentDeposition} result in an enhancement of the energy threshold for oxidation. For this reason the oxidation process for thicker MoS$_{2}$ configurations is less pronounced under the chosen treatment parameters, what is also observed with other analytic methods \cite{zhou2013thickness}. Furthermore, the calculations suggest, that the modifications of the band structure by oxidation are more pronounced for 1L and 2L, while showing weaker changes for higher layer numbers.
For the 1L, the calculation of the band structure plotted in Figure \ref{fig:bandstructure}, shows a transition from a direct to an indirect band gap for a full oxidation of the top most sulfur layer, resulting in a pronounced change in the SH-intensity. Furthermore, the 1L behavior is influenced by the chosen substrate, hence its direct contact to them \cite{Yang2024IsolatingSubstrate}. For SiO$_{2}$, the 1L is more stable than a higher number of layers, due to a higher energy barrier for the MoS$_{2}$-substrate interface, in comparison to the MoS$_{2}$-MoS$_{2}$ multilayer interface \cite{Wang2018SubstrateMoS2}. However, the high surface-to-volume ratio of the 1L is a counteract to this behavior, what becomes relevant for high defect rates like S vacancies at the surface and/or sufficient high oxidation temperature, which allow overcoming the energy barrier introduced by the substrate.

\section{Summary}

The thermal oxidation process of atomically thin \ce{MoS2} is investigated utilizing SH-microscopy. For this purpose, \ce{MoS2} flakes have been fabricated by mechanical exfoliation and thermally oxidized in controlled O-rich environment for different times. The samples are inspected by optical microscopy and the layer thickness is determined using Raman spectroscopy, which reveals MoS$_{2}$ flakes ranging from 1L to 7L. In a first nonlinear analysis, the fundamental layer dependent nonlinear behavior for untreated and oxidized samples is compared. Here, for odd numbered MoS$_{2}$, and besides the typical thickness dependency, a specific SH-decrease is found. This is caused by oxidation attributed to distinct changes in the band structure combined with a reduction of the effective crystal available for SHG. For even numbered structures, which show only a residual signal, oxidation leads to an enhancement of the SH-signal due to a breaking of symmetry. Additionally, calculations of the band structure of 1L -3L \ce{MoS2} for a full exchange of the top S layer with O atoms have been carried out, showing significant modulation with the oxidation, especially for the 1L, where the band gap transitions from direct to indirect. Moreover, the nonlinear behavior with respect to the oxidation time ($t_a=0-6$h) in excitation power dependent measurements is analyzed. The ongoing decrease of the SH-intensity for odd number of layers (and increase for even number of layers) with increasing oxidation time indicates a progressive oxidation within the maximum oxidation time of six hours and a full oxidation of the top-most \ce{MoS2} layer can be excluded. For odd number of layers, a polarization dependent analysis reveals the six-fold dependency of the SH-intensity. With oxidation, the difference between maximum and minimum intensity decreases, while the symmetry is retained. Non-oxidized even numbered layers give no angle depend polarization intensity and only the residual signal is detected. Upon oxidation, a six-fold symmetry arises, what is evidence of the breaking of inversion symmetry by thermal oxidation. Furthermore, a thickness dependent oxidation behavior is observed. The lower the layer number, the more O atoms are introduced in the crystal, with the monolayer representing a special case due to its direct contact with the substrate.

In conclusion, the study of 1L up to 7L \ce{MoS2} demonstrates, that the thermal oxidation progress can be monitored with non-resonant and non-invasive SH microscopy thanks to distinct fingerprints of structural modification in the nonlinear response and reveals a layer number dependent oxidation process, whose understanding is fundamental for the design and realization of \ce{MoS2}-based devices.

\section{Acknowledgments}
This work is supported by the Deutsche Forschungsgemeinschaft (German Research Foundation) through the transregional collaborative research center TRR142/3-2022 and by the European Research Council (ERC) under the European Union’s Horizon 2020 research and innovation program (LiNQs, 101042672). K.D.J. acknowledges funding from the Ministry of Culture and Science of North Rhine-Westphalia for the Institute for Photonic Quantum Systems (PhoQS).
We also thank the Paderborn Center for Parallel Computing (PC$^2$) for grants of high-performance computational time. K.B. thanks R. Adhikari for the fruitful discussions and his insights.

\bibliography{bibliography}

\end{document}